  \documentclass[conference]{IEEEtran}
 \normalsize
\usepackage{algorithm,algpseudocode}
\usepackage{algorithm}
\usepackage{epsfig}
\usepackage{physics}
\usepackage{amsmath}
\usepackage{amsmath}
\usepackage{theorem}
\usepackage{amsmath}
\usepackage{mathrsfs}
\usepackage{subcaption}
\usepackage{booktabs}
\usepackage{pgfplots}
\usepackage{grffile}
\usepackage{epsfig}
\usepackage{amsfonts}
\usepackage{amssymb}
\usepackage{tabularx}
\usepackage{mathrsfs}
\usepackage{euscript}
\usepackage{graphicx}
\usepackage{multirow}
\usepackage{stfloats}
\usepackage{cite,stix}
\usepackage{subcaption}
\usepackage[T1]{fontenc}
\usepackage{booktabs}
\usepackage{pgfplots}
\usepackage{grffile}
\pgfplotsset{compat=newest}
\usetikzlibrary{plotmarks}
\usetikzlibrary{arrows.meta}
\usepgfplotslibrary{patchplots}
\usepackage{amsmath,color}
\newcommand{\E}{\mathbb E }
\usepackage{color}
\usepackage{subcaption}

\begin{document}
	\title{  Full-Duplex GFDM Radio Transceivers in the Presence of Phase Noise, CFO and  IQ Imbalance }

    \author{\IEEEauthorblockN{Amirhossein Mohammadian, \textit{Student Member, IEEE}, and Chintha Tellambura, \textit{Fellow, IEEE}}
\IEEEauthorblockA{Department of Electrical and Computer Engineering
University of Alberta, Edmonton, Alberta, Canada\\
Email: am11@ualberta.ca and chintha@ece.ualberta.ca
}}
\markboth{}%
{}	
	\maketitle	
	\begin{abstract}
This paper addresses the performance of a  full-duplex (FD) generalized frequency division multiplexing (GFDM) transceiver in the presence of radio frequency (RF) impairments including  phase noise, carrier frequency offset (CFO) and in-phase (I) and quadrature (Q) imbalance. We study analog and digital self-interference (SI) cancellation  and develop a  complementary SI suppression method. Closed-form solutions  for the residual SI power and the desired signal power and signal-to-interference ratio (SIR) are  provided. Simulation results show that the RF impairments degrade   SI cancellation  and FD GFDM is more sensitive to them compares to  FD orthogonal frequency division multiplexing (OFDM). Hence, we propose an FD GFDM  receiver filter for maximizing the SIR. Significantly,   it  achieves 25\:dB higher SIR  than  FD OFDM transceiver.         
\end{abstract}


	\IEEEpeerreviewmaketitle
	
\section{Introduction}\label{sec:intro}

Due to the  increasing wireless data and emerging  fifth generation (5G) networks, full-duplex (FD) radio may be the answer to  handle higher capacity demands  \cite{magh1, FD1}. FD radios can  simultaneously transmit and receive on the same frequency and  time slots,  which potentially doubles the capacity, reduces network delay, improves network secrecy and increases spectrum usage flexibility \cite{FD1,FD2}.  Applications of FD wireless are numerous (see \cite{mohammadi2015} and \cite{mohammadi2018} and references therein).  Moreover, 4G wireless  cellular deploys  orthogonal frequency division multiplexing (OFDM), which may not be sufficient to  reach  all vital requirements of 5G. A potential alternative to OFDM is generalized frequency division multiplexing (GFDM),  a filtered multicarrier modulation scheme with low out-of-band (OOB)  emissions, high spectral efficiency and low latency \cite{GFDMo,GFDM-PA-Kh}. Thus, the combination of  GFDM  and FD may  be the ideal architecture to achieve  5G network requirements.

The fundamental  challenge in the FD radios  is the self-interference (SI) due to the coupling  of  the  transmit signal to the receiver path during simultaneous transmission and reception.  To mitigate SI,  active cancellation  is performed over the analog and digital parts of the receiver chain \cite{FD-cancel}. In the analog part, the dominant SI component is suppressed by subtracting adjusted transmitted signal in amplitude, time and phase from the received signal. The rest of the multipath components are processed in digital part by estimating the channel-state information (CSI). However, in direct conversion transceivers, radio frequency (RF) impairments in front-end components including phase noise,  carrier frequency offset (CFO) and in-phase (I) and quadrature (Q) imbalance degrade the link performance significantly. In the FD system, these RF imperfections will limit the capability of the SI cancellation mechanisms,  which must  be carefully considered when evaluating the system performance. 

In \cite{FD-PN}, the authors investigate the harmful effects of phase noise on the SI cancellation capability of an FD OFDM  transceiver and observe that phase noise limits the performance of the SI suppression techniques. The same transceiver under nonlinear power amplifier and IQ imbalance is studied in \cite{FD-PA}. It is shown that IQ imbalance adds image components to the SI signal which have detrimental effects on SI cancellation. Moreover, in \cite{FD-let}, phase noise and IQ imbalance in an FD OFDM transceiver are investigated. A scheme for estimation and cancellation the effects of the IQ imbalance, power amplifier nonlinearity and phase noise on an FD OFDM transceiver is proposed in \cite{FD-inter}. For the GFDM waveform, impacts of timing offset, CFO and phase noise are studied in \cite{SIR-GFDM}. Optimal filter design for a GFDM transceiver in presence of CFO is presented in \cite{GFDM-filter}.  GFDM performance in cognitive radio is studied in \cite{GFDM1, GFDM2} and effects of the nonlinear power amplifier are investigated for that. Digital interference cancellation scheme for an FD GFDM transceiver is proposed in \cite{FD-GFDM} and SI power is calculated as well. Analog cancellation which suppresses the dominant part of SI power is not considered in \cite{FD-GFDM} and the effects of the RF impairments are not analyzed. Furthermore, to our best knowledge, the FD GFDM transceiver has not been properly modeled with analog and digital SI cancellation and  impacts of the phase noise, CFO and IQ imbalance have not been studied.  In this paper, we study the FD GFDM transceiver performance in presence of the phase noise, CFO and IQ imbalance. 
 
 
 In detail, the contributions of this paper are as follows:
 \begin{itemize}
	\item We fully  model the FD GFDM transceiver by considering the phase noise, CFO and IQ imbalance. Both analog and digital SI cancellation stages are addressed and a complementary method for more suppression of the SI signal in digital domain is developed.
	\item Residual SI power after analog and digital SI cancellations is derived. Moreover, power of the intended signal in presence of the RF impairments in the receiver is presented. To the best of our knowledge, the collective impact of phase noise, CFO and IQ imbalance  has not been investigated for GFDM half-duplex (HF) transceivers,  which we do in this paper.
	\item By utilizing the derived signal, signal-to-interference ratio (SIR) for received signal is derived. Based on our analysis, we find that FD GFDM is more sensitive to the RF impairments compared to  FD OFDM. To mitigate  this problem, we design an optimal FD-GFDM receiver filter  to maximize the SIR.
	\item All the theoretical derivations are verified with simulation results. Moreover, to determine the performance gains of  FD GFDM transceiver, we also present  FD OFDM transceiver results.
	
\end{itemize}
	\section{System Model}\label{sec:model}
    
The signal model of FD GFDM transceiver in presence of RF impairments including  phase noise, CFO and IQ imbalance is shown in Fig. 1.  The  transmitter and the receiver deploy single separate antennas and the  well-known direct-conversion architecture.   The SI cancellation relies on   analog linear cancellation (ALC) in first stage of the receiver and digital linear cancellation (DLC) in baseband unit.  In following, we model and analyze the system in detail.

The GFDM transmitter generates  the signal in which data of $M$-th time-slots are transmitted on the $K$-th subcarriers. For one symbol time, the GFDM signal  may  be expressed as


	\begin{figure}[t]
	\centering
		\includegraphics[width=.5\textwidth]{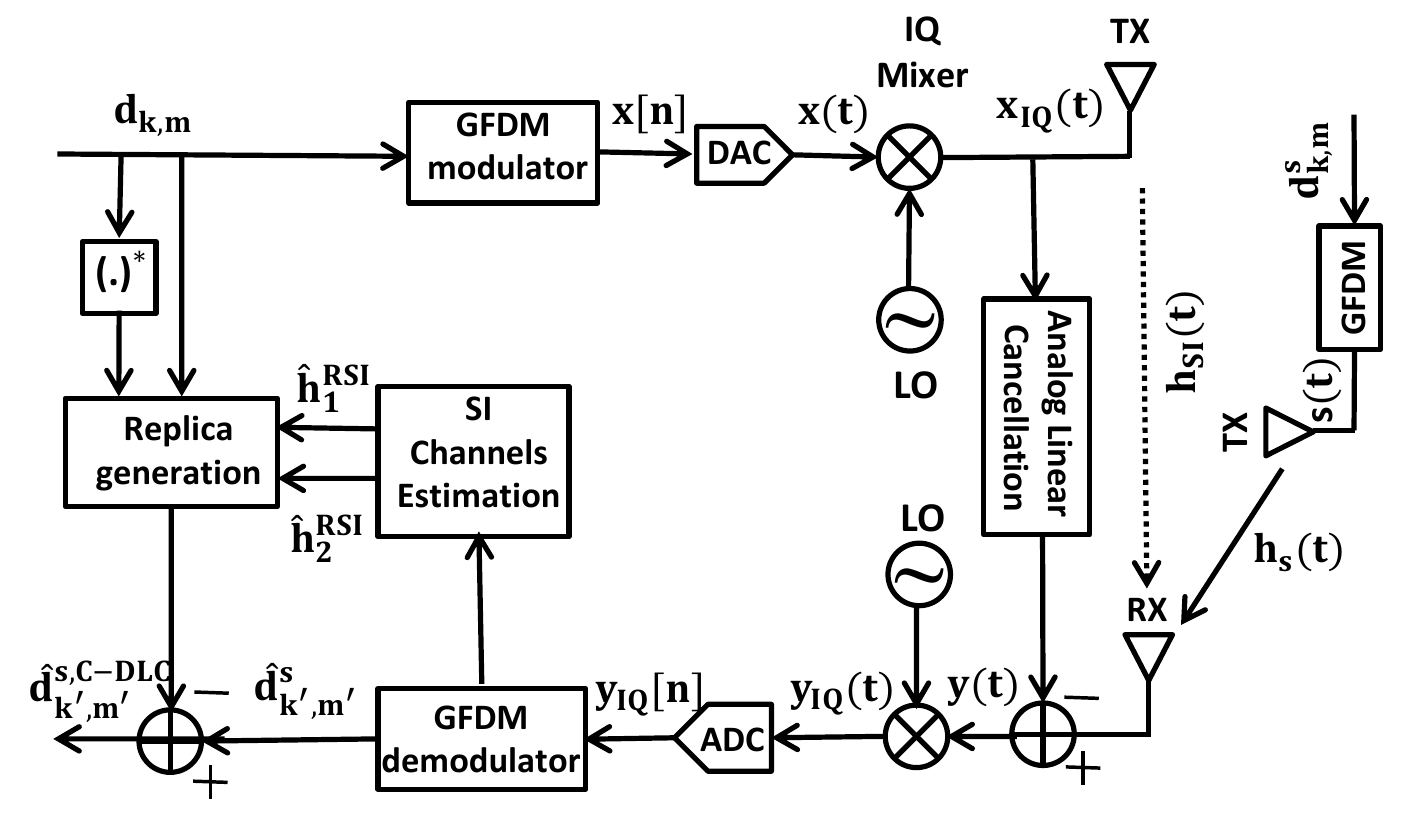}\\
		\caption {FD GFDM transceiver. }
		\label{system model}
	\end{figure}
 \begin{equation}\label{equ1}
  x[n]=\sum\limits_{k=0}^{K-1}{\sum\limits_{m=0}^{M-1}{{{d}_{k,m}}}{{g}   _  {m}}[n]{{e}^{\frac{j2\pi kn}{K}}}},~~~0\leq n\leq MK-1 
	\end{equation}
where ${{d}_{k,m}}$ is zero mean  independent and identically distributed (i.i.d.) complex data symbol on $k$-th subcarrier of $m$-th time-slot with symbol energy $P_{d}$ and ${{g}_{m}}[n]=g{{[n-mK]}_{MK}}$ is a circularly shifted version of normalized  prototype filter $g[n]$ ($\sum\limits_{n=0}^{MK-1}{|g[n]|}^{2}=1$).  With the addition of  cyclic prefix (CP) and passing through digital-to-analog conversion (DAC), the  analog baseband signal, $x(t)$, is passed through IQ mixer. Due to  mismatches between the amplitudes and phases of I-and Q-branches,  an undesired signal,  which is the  mirror image of the original signal, is added. Thus, the  output signal of the IQ mixer may  be written as 
 \begin{equation}\label{equ2}
x_{IQ}(t)=(g_{Tx,d}x(t)+g_{Tx,I}x^{*}(t))e^{j\phi_{Tx}(t)}
	\end{equation}
where $(.)^*$ indicates complex conjugate, $g_{Tx,d}$ and $g_{Tx,I}$ are the transmitter IQ mixer responses for the direct and image signals, respectively, and $\phi_{Tx}(t)$ is random phase noise of the local oscillator of the transmitter side. After passing the transmitter signal through wireless channel and applying  ALC, the received signal could be formulated as

 \begin{equation}\label{equ3}
y(t)=s(t)*h_s(t)+x_{IQ}(t)*h_{RSI}(t)+w(t)
	\end{equation}
where $*$ denotes the convolution, $s(t)$ is desired signal, $h_s(t)$ is multipath desired channel from the intended transmitter to the local receiver, $h_{RSI}(t)=h_{SI}(t)-h_{ALC}(t)$ is residual SI channel where $h_{ALC}(t)$ is estimate of the the multipath coupling  channel and $h_{SI}(t)$ is the multipath coupling  channel between the local transmitter and the receiver, and $w(t)$ is a Gaussian noise. Then, the signal goes through the receiver IQ mixer which the output is written as
 \begin{equation}\label{equ5}
y_{IQ}(t)=g_{Rx,d}y(t)e^{-j\phi_{Rx}(t)}e^{j2 \pi \Delta_{f} t}+g_{Rx,I}y^{*}(t)e^{j\phi_{Rx}(t)}e^{-j2 \pi \Delta_{f} t}
	\end{equation}
where $g_{Rx,d}$ and $g_{Rx,I}$ are the receiver IQ mixer responses for the direct and image signals. $\phi_{Rx}(t)$ is random phase noise of the local oscillator of the receiver side and $\Delta_{f}$ indicates the difference between carrier frequency of the receiver and transmitter local oscillators. Image rejection ratio ($\text{IRR}$) is considered for quantifying the quality of the IQ mixer which is defined as $\text{IRR}_{Rx}=\frac{|g_{Rx,I}|^2}{|g_{Rx,d}|^2}$.
By assuming $L$-tap propagation channels $(h[n]=\sum_{l=0}^{L-1}h_{l}\delta[n-l])$, according to (\ref{equ2}), (\ref{equ3}) and (\ref{equ5}), the discrete sample of the signal could be expressed as
 \begin{equation}\label{equ6}
 \begin{aligned}
y_{IQ}[n]=&\sum_{l=0}^{L-1}h^{RSI}_{1}[n,l]x[n-l]+h^{RSI}_{2}[n,l]x^{*}[n-l]+\\& h^{s}_{1}[n,l]s[n-l]+h^{s}_{2}[n,l]s^{*}[n-l]+w_d[n]+w_{I}[n]
\end{aligned}
	\end{equation}
where equivalent channel responses for individual signal components can be written as
 \begin{equation}\label{equ7}
 \begin{aligned}
h^{RSI}_{1}[n,l]=&g_{Tx,d}g_{Rx,d}h_{RSI,l}e^{j(\phi_{TX}[n-l]-\phi_{RX}[n])}e^{\frac{j2\pi\epsilon n}{K}}+\\
&g_{Tx,I}^{*}g_{Rx,I}h_{RSI,l}^{*}e^{-j(\phi_{TX}[n-l]-\phi_{RX}[n])}e^{\frac{-j2\pi\epsilon n}{K}}\\
h^{RSI}_{2}[n,l]=&g_{Tx,I}g_{Rx,d}h_{RSI,l}e^{j(\phi_{TX}[n-l]-\phi_{RX}[n])}e^{\frac{j2\pi\epsilon n}{K}}+\\
&g_{Tx,d}^{*}g_{Rx,I}h_{RSI,l}^{*}e^{-j(\phi_{TX}[n-l]-\phi_{RX}[n])}e^{\frac{-j2\pi\epsilon n}{K}}\\
h^{s}_{1}[n,l]=&g_{Rx,d}h_{s,l}e^{-j\phi_{RX}[n]}e^{\frac{j2\pi\epsilon n}{K}}\\
h^{s}_{2}[n,l]=&g_{Rx,I}h^{*}_{s,l}e^{j\phi_{RX}[n]}e^{\frac{-j2\pi\epsilon n}{K}}\\
w_d[n]=&g_{Rx,d}e^{-j\phi_{RX}[n]}e^{\frac{j2\pi\epsilon n}{K}}w[n]\\
w_d[n]=&g_{Rx,I}e^{j\phi_{RX}[n]}e^{\frac{-j2\pi\epsilon n}{K}}w^{*}[n]
\end{aligned}
\end{equation}
where $\epsilon$ is normalized CFO. At the  digital domain before deploying DLC, samples are sent to GFDM demodulator where the estimated symbol  at  ${{k}^{'}}$-th subcarrier  and ${{m}^{'}}$-th time-slot  is 
\begin{equation}\label{equ8}
{{\widehat{d}}_{{{k}^{'}},{{m}^{'}}}}^{s}=\sum\limits_{n=0}^{MK-1}{(y_{IQ}[n])f_{{{{m}^{'}}}}[n]{{e}^{ \frac{-j2\pi k^{'}n}{K}}}}
	\end{equation}    
where ${{f}_{_{m}}}[n]=f{{[n-mK]}_{MK}}$ is circularly shifted version of  receiver filter impulse response $f[n]$. Finally, to further decrease the residual SI signal, by utilizing the replica of transmitted symbols and estimation of the equivalent residual SI channels, digital cancellation symbols are generated and subtracted from the demodulated symbols which is named by classical DLC. Furthermore,\cite{FD-PA} shows that after the classical DLC, the image of the SI signal is the dominant source of distortion. Thus, complementary DLC (C-DLC) is proposed in which the cancellation of the image of SI signal is done similar to classical DLC by deploying the conjugate of the transmitted symbols. The output of C-DLC could be expressed as

 \begin{equation}\label{equ9}
 \begin{aligned}
{{\widehat{d}}^{s,C-DLC}_{{{k}^{'}},{{m}^{'}}}}=&\left(R^{SI}_{k',m'}-R^{DLC}_{k',m'}\right)+\left(R^{SI,im}_{k',m'}-R^{DLC,i}_{k',m'}\right)+\\&R^{s}_{k',m'}+R^{s,im}_{k',m'}+w^{eq}_{k',m'}+w^{eq,im}_{k',m'}
\end{aligned}
	\end{equation}
where $R^{SI}_{k',m'}$, $R^{SI, im}_{k',m'}$, $R^{s}_{k',m'}$, $R^{s,im}_{k',m'}$, $w^{eq}_{k',m'}$ and $w^{eq, im}_{k',m'}$  are corresponding terms for SI signal, intended signal and the equivalent noise after GFDM demodulator that are derived from (\ref{equ1}) and (\ref{equ6})-(\ref{equ8}). Moreover, $R^{DLC}_{k',m'}$ and $R^{DLC,i}_{k',m'}$ are DLC terms for linear and conjugate replica of the symbol, which are written as

 \begin{equation}\label{equ10}
 \begin{aligned}
R^{DLC}_{k',m'}=&{{{d}_{k',m'}}}\sum_{l=0}^{L-1}\sum_{n=0}^{MK-1}\hat h^{RSI}_{1}[n,l]f_{{{{m}^{'}}}}[n]{{g}   _  {m'}}[n-l]{{e}^{ \frac{-j2\pi{k^{'}}l}{K}}}\\
R^{DLC,i}_{k',m'}=&d^{*}_{k',m'}\sum_{l=0}^{L-1}\sum_{n=0}^{MK-1}\hat h^{RSI}_{2}[n,l]f_{{{{m}^{'}}}}[n]g^{*}_{m'}[n-l]{{e}^{ \frac{-j2\pi k^{'}(2n-l)}{K}}}\\
\end{aligned}
\end{equation}
where $\hat h^{RSI}_{1}[n,l]$ and $\hat h^{RSI}_{2}[n,l]$ indicate equivalent channel estimation of the linear SI signal and the conjugate SI signal, respectively. It is worth mentioning that output of the classical DLC is derived by, ${{\widehat{d}}^{s,DLC}_{{{k}^{'}},{{m}^{'}}}}={{\widehat{d}}^{s,C-DLC}_{{{k}^{'}},{{m}^{'}}}}+R^{DLC,i}_{k',m'}$

\section{Signal power analysis}
In this section, we calculate the power of the residual SI signal and the desired signal in closed-form. The channels, transmitted data and phase noise are assumed as independent random processes. Furthermore, perfect channel estimation and two independent oscillators for the local transmitter and the receiver are considered in this paper.

\subsection{SI signal power analysis}
According to wide-sense stationary uncorrelated scattering (WSSUS) model, $\forall l$ : $h_{RSI,l}$ are assumed independent of each other, $\E [h_{RSI,l}]=0$ and $\E \left [\abs{h_{RSI,l}}^{2}\right ]=\sigma^{2}_{RSI,l}$. $\E[.]$ indicates the statistical expectation operator.  Furthermore, free-running oscillators (FRO) with  Brownian motion process are used for generating the phase noise $[\phi[n+1]-\phi[n]]\sim \mathcal{N} (0,4\pi\beta T_{s})$, where $\phi[n]$ is Brownian motion with 3-dB bandwidth of $\beta $ and $T_{s}$ is the sample interval. Accordingly, after straight-forward manipulation, variance of the linear residual SI After ALC, $\sigma^{SI-ALC}_{k',m'}=\E \left[\abs{R^{SI}_{k',m'}}^{2} \right ]$, is derived as
 \begin{equation}\label{equ12}
 \begin{aligned}
&\sigma^{SI-ALC}_{k',m'}= P_{d} \sum_{n_{1}=0}^{MK-1}\sum_{n_{2}=0}^{MK-1}f_{m'}[n_{1}]f^{*}_{m'}[n_{2}]e^{-4\abs{n_{1}-n_{2}}\pi \beta T_{s}}\\&\bigg(\abs{g_{TX,d}g_{RX,d}}^2e^{\frac{j2\pi (n_{1}-n_{2})\epsilon }{K}}+\abs{g_{TX,I}g_{RX,I}}^2e^{\frac{-j2\pi (n_{1}-n_{2})\epsilon }{K}}\bigg)\\\times &\sum_{l=0}^{L-1}\sum_{k=0}^{K-1}\sum_{m=0}^{M-1} \sigma^{2}_{RSI,l}   g_{m}[n_{1}-l]g^{*}_{m}[n_{2}-l]e^{\frac{j2\pi (n_{1}-n_{2})(k-k^{'}) }{K}}.
\end{aligned}
	\end{equation}
	
On the other hand, the power of the linear residual SI  after C-DLC can be defined as $\sigma^{SI-DLC}_{k',m'}=\E \left[\abs{R^{SI}_{k',m'}-R^{DLC}_{k',m'}}^{2} \right ]$ which is given by 

 \begin{equation}\label{equ13}
 \begin{aligned}
&\sigma^{SI-DLC}_{k',m'}= P_{d} \sum_{n_{1}=0}^{MK-1}\sum_{n_{2}=0}^{MK-1}f_{m'}[n_{1}]f^{*}_{m'}[n_{2}]e^{-4\abs{n_{1}-n_{2}}\pi \beta T_{s}}\\&\bigg(\abs{g_{TX,d}g_{RX,d}}^2e^{\frac{j2\pi (n_{1}-n_{2})\epsilon }{K}}+\abs{g_{TX,I}g_{RX,I}}^2e^{\frac{-j2\pi (n_{1}-n_{2})\epsilon }{K}}\bigg)\\&\times   \sum_{l=0}^{L-1}\underset{k\neq k' \& m \neq m'}{\sum_{k=0}^{K-1}\sum_{m=0}^{M-1}} \sigma^{2}_{RSI,l}   g_{m}[n_{1}-l]g^{*}_{m}[n_{2}-l]e^{\frac{j2\pi (n_{1}-n_{2})(k-k^{'}) }{K}}.
\end{aligned}
	\end{equation}

Note that  (\ref{equ12}) and (\ref{equ13}) depend on  multipath profile, 3-dB phase noise bandwidth, normalized CFO, IQ imbalance coefficients, number of subcarriers and time-slots and GFDM receiver and transmitter filters. Thus, all these parameters affect  the efficiency of analog and digital SI cancellations. Similarly, the power of the conjugate residual SI after ALC and after C-DLC could be formulated as 
 \begin{equation}\label{equ14}
 \begin{aligned}
&\sigma^{SI-im-ALC}_{k',m'}= P_{d} \sum_{n_{1}=0}^{MK-1}\sum_{n_{2}=0}^{MK-1}f_{m'}[n_{1}]f^{*}_{m'}[n_{2}]e^{-4\abs{n_{1}-n_{2}}\pi \beta T_{s}}\\&\left(\abs{g_{TX,I}g_{RX,d}}^2e^{\frac{j2\pi (n_{1}-n_{2})\epsilon }{K}}+\abs{g_{TX,d}g_{RX,I}}^2e^{\frac{-j2\pi (n_{1}-n_{2})\epsilon }{K}}\right)\\\times &\sum_{l=0}^{L-1}\sum_{k=0}^{K-1}\sum_{m=0}^{M-1} \sigma^{2}_{RSI,l}   g^{*}_{m}[n_{1}-l]g_{m}[n_{2}-l]e^{\frac{-j2\pi (n_{1}-n_{2})(k+k^{'}) }{K}}
\end{aligned}
	\end{equation}
and

 \begin{equation}\label{equ15}
 \begin{aligned}
&\sigma^{SI-im-DLC}_{k',m'}= P_{d} \sum_{n_{1}=0}^{MK-1}\sum_{n_{2}=0}^{MK-1}f_{m'}[n_{1}]f^{*}_{m'}[n_{2}]e^{-4\abs{n_{1}-n_{2}}\pi \beta T_{s}}\\&\left(\abs{g_{TX,I}g_{RX,d}}^2e^{\frac{j2\pi (n_{1}-n_{2})\epsilon }{K}}+\abs{g_{TX,d}g_{RX,I}}^2e^{\frac{-j2\pi (n_{1}-n_{2})\epsilon }{K}}\right)\\&\times   \sum_{l=0}^{L-1}\underset{k\neq k' \& m \neq m'}{\sum_{k=0}^{K-1}\sum_{m=0}^{M-1}} \sigma^{2}_{RSI,l}   g^{*}_{m}[n_{1}-l]g_{m}[n_{2}-l]e^{\frac{-j2\pi (n_{1}-n_{2})(k+k^{'}) }{K}}
\end{aligned}
\end{equation}
where $\sigma^{SI-im-ALC}_{k',m'}=\E \left[\abs{R^{SI,im}_{k',m'}}^{2} \right ]$ and $\sigma^{SI-im-DLC}_{k',m'}=\E \left[\abs{R^{SI,im}_{k',m'}-R^{DLC,i}_{k',m'}}^{2} \right ]$. Again, the results depend on the system parameters and the performance of the system can be evaluated for different configurations. Following (\ref{equ13}) and (\ref{equ15}), total power of residual SI signal after C-DLC may be expressed as
 \begin{equation}\label{equ16}
 \sigma^{SI}_{k',m'}= \sigma^{SI-DLC}_{k',m'}+ \sigma^{SI-im-DLC}_{k',m'}.
	\end{equation}

\subsection{Desired signal power analysis }

We assume that the desired signal is generated by (\ref{equ1}) with i.i.d input symbols of  ${{d}^{s}_{k,m}}$ with symbol energy $p_{d}$. No imperfections are considered in the transmitter of the desired signal. Thus, the desired symbol could be extracted from $R^{s}_{k',m'}$ as

\begin{equation}\label{equ18}
d^{ss}_{k',m'}=d^{s}_{k',m'}\sum_{l=0}^{L-1}\sum_{n=0}^{MK-1}h^{s}_{1}[n,l]f_{{{{m}^{'}}}}[n]g_{m'}[n-l]{{e}^{\frac{{-j2\pi {k}^{'}}l}{K}}}.
\end{equation}

Following the WSSUS model, $\forall l$ : $h_{s,l}$ are assumed to be independent of each other, $\E [h_{s,l}]=0$ and $\E \left [\abs{h_{s,l}}^{2}\right ]=\sigma^{2}_{s,l}$. Therefore, the variance of the desired symbol could be derived by
\begin{equation}\label{equ19}
\begin{aligned}
&\sigma^{s}_{k',m'}=\E \left[\abs{d^{ss}_{k',m'}}^{2} \right ]=\abs{g_{RX,d}}^2 P_{d} \sum_{l=0}^{L-1}\sum_{n_{1}=0}^{MK-1}\sum_{n_{2}=0}^{MK-1} \sigma^{2}_{s,l}\\& e^{-2\abs{n_{1}-n_{2}}\pi \beta T_{s}} f_{m'}[n_{1}]f^{*}_{m'}[n_{2}]g_{m'}[n_{1}-l]g^{*}_{m'}[n_{2}-l]e^{\frac{j2\pi (n_{1}-n_{2})\epsilon }{K}}.
\end{aligned}
\end{equation}

According to the desired symbol, interference signals could be considered as $R^{s}_{k',m'}-d^{ss}_{k',m'}$ and $R^{s,im}_{k',m'}$. The variance of the first term could be calculated as $\sigma^{R^{s}}_{k',m'}-\sigma^{s}_{k',m'}$ where $\sigma^{R^{s}}_{k',m'}=\E \left[\abs{R^{s}_{k',m'}}^{2} \right ]$ is equal to 

\begin{equation}\label{equ20}
 \begin{aligned}
\sigma^{R^{s}}_{k',m'}&= \abs{g_{RX,d}}^2 P_{d} \sum_{l=0}^{L-1}\sum_{n_{1}=0}^{MK-1}\sum_{n_{2}=0}^{MK-1} \sum_{k=0}^{K-1}\sum_{m=0}^{M-1}\sigma^{2}_{s,l}e^{-2\abs{n_{1}-n_{2}}\pi \beta T_{s}}\\&f_{m'}[n_{1}]f^{*}_{m'}[n_{2}]  g_{m}[n_{1}-l]g^{*}_{m}[n_{2}-l]e^{\frac{j2\pi (n_{1}-n_{2})(\epsilon+k-k^{'}) }{K}}.
\end{aligned}
\end{equation}

Moreover, the variance of the second term could be expressed as

 \begin{equation}\label{equ21}
 \begin{aligned}
\sigma^{R^{s,im}}_{k',m'}&= \abs{g_{RX,I}}^2 P_{d}  \sum_{l=0}^{L-1}\sum_{n_{1}=0}^{MK-1}\sum_{n_{2}=0}^{MK-1} \sum_{k=0}^{K-1}\sum_{m=0}^{M-1}\sigma^{2}_{s,l}e^{-2\abs{n_{1}-n_{2}}\pi \beta T_{s}}\\&f_{m'}[n_{1}]f^{*}_{m'}[n_{2}]  g^{*}_{m}[n_{1}-l]g_{m}[n_{2}-l]e^{\frac{-j2\pi (n_{1}-n_{2})(\epsilon+k+k^{'}) }{K}}.
\end{aligned}
	\end{equation}
	
Thus, the total power of the interference signal is given by

 \begin{equation}\label{equ22}
 \sigma^{s,i}_{k',m'}=\sigma^{R^{s}}_{k',m'}+\sigma^{R^{s,im}}_{k',m'}-\sigma^{s}_{k',m'}.
 \end{equation}

\section{SIR formulation and filter optimization}

Herein, the SIR of the FD GFDM transceiver is derived and a receiver filter for maximizing the SIR is proposed. According to (\ref{equ16}), (\ref{equ19}) and (\ref{equ22}), SIR of the estimated symbol in $k'$-th subcarrier and $m'$-th subsymbol  may be expressed as 
 \begin{equation}\label{equ26}
 \Gamma_{k',m'}=\frac{\sigma^{s}_{k',m'}}{\sigma^{SI}_{k',m'}+\sigma^{s,i}_{k',m'}}. 
 \end{equation}
 
Since GFDM use  non-orthogonal subcarriers, it performs worse than  OFDM in the  presence of RF impairments. Thus,  FD GFDM should achieve lower SIR  than FD OFDM. However, GFDM contains degrees of freedom  in receive  filter design that can help us to improve the performance.  To retain the benefits  of GFDM such as lower  out-of-band  emissions,  conventional filter is assumed for the transmitter side.  On the other hand, the  receiver filter is optimized to maximize the SIR. Let us denote $\mathbf {f_{k',m'}}=\mathbf {S}_{k'}\mathbf {M}_{m'M}\mathbf {f}_{0,0}\in \mathbb{C}^{MK \times 1}$ contains samples of $f_{k',m'}[n]=f_{{{{m}^{'}}}}[n]{{e}^{ \frac{-j2\pi k^{'} n}{K}}}$ in (\ref{equ8}) where $\mathbf{f}_{0,0}\in \mathbb{C}^{MK \times 1}$ is the column vector including receiver filter $f[n]$ samples, $\mathbf{M}_{m'M}\in \mathbb{C}^{MK \times MK}$ circularly shifts $\mathbf{f}_{0,0}$  and $\mathbf{S}_{k'}=\text{diag}\left(\left[1, e^{\frac{-j2 \pi k'}{K}}, ..., e^{\frac{-j2 \pi k'(MK-1)}{K}}\right]\right)\in \mathbb{C}^{MK \times MK}$ is the subcarrier mapping matrix. It is worth mentioning that (\ref{equ8}) could be expressed as $\hat {d}^{s}_{k',m'}={{\mathbf{y}}}_{IQ}\mathbf{f}_{k',m'}$ where ${{\mathbf{y}}}_{IQ} \in \mathbb{C}^{1 \times MK}$ contains ${y}_{IQ}[n]$. Moreover, according to derivations, we rewrite the derived variances in matrix form as $\sigma^{s}_{k',m'}={\mathbf{f}^{H}_{k',m'}\mathbf{U}_{m'}\mathbf{f}_{k',m'}}$,  $\sigma^{SI}_{k',m'}= {\mathbf{f}^{H}_{k',m'}\mathbf{V}^{SI}\mathbf{f}_{k',m'}}$ and  $\sigma^{s,i}_{k',m'}=\left({\mathbf{f}^{H}_{k',m'}\mathbf{V}^{R}\mathbf{f}_{k',m'}}-{\mathbf{f}^{H}_{k',m'}\mathbf{U}_{m'}\mathbf{f}_{k',m'}}\right)$  where

     \begin{equation}\label{equ227}
 \begin{aligned}
U_{m'}[n_2,n_1]=&\sum_{l=0}^{L-1}\abs{g_{RX,d}}^2 P_{d}  \sigma^{2}_{s,l}e^{-2\abs{n_{1}-n_{2}}\pi \beta T_{s}} g_{m'}[n_{1}-l]\\&g^{*}_{m'}[n_{2}-l]e^{\frac{j2\pi (n_{1}-n_{2})\epsilon }{K}}.
\end{aligned}
	\end{equation}
and
    \begin{equation}\label{equ28}
 \begin{aligned}
&V^{SI}[n_2,n_1]= \sum_{l=0}^{L-1}\underset{k\neq k' \& m \neq m'}{\sum_{k=0}^{K-1}\sum_{m=0}^{M-1}}  P_{d} \sigma^{2}_{RSI,l}  g_{m}[n_{1}-l]g^{*}_{m}[n_{2}-l]\\&e^{-4\abs{n_{1}-n_{2}}\pi \beta T_{s}}\Big(\abs{g_{TX,d}g_{RX,d}}^2e^{\frac{j2\pi (n_{1}-n_{2})(\epsilon+k) }{K}}+\abs{g_{TX,I}g_{RX,I}}^2\\&e^{\frac{-j2\pi (n_{1}-n_{2})(\epsilon-k) }{K}}+\abs{g_{TX,I}g_{RX,d}}^2e^{\frac{j2\pi (n_{1}-n_{2})(\epsilon-k) }{K}}+\abs{g_{TX,d}g_{RX,I}}^2\\&e^{\frac{-j2\pi (n_{1}-n_{2})(\epsilon+k) }{K}}\Big).
\end{aligned}
	\end{equation}
and    
    \begin{equation}\label{equ29}
 \begin{aligned}
&V^{R}[n_2,n_1]=\sum_{l=0}^{L-1}\sum_{k=0}^{K-1}\sum_{m=0}^{M-1} P_{d} \sigma^{2}_{s,l} g_{m}[n_{1}-l]g^{*}_{m}[n_{2}-l]\\&e^{-2\abs{n_{1}-n_{2}}\pi \beta T_{s}}\Big(\abs{g_{RX,d}}^2e^{\frac{j2\pi (n_{1}-n_{2})(\epsilon+k) }{K}}+\abs{g_{RX,I}}^2e^{\frac{-j2\pi (n_{1}-n_{2})(\epsilon+k) }{K}}\Big).
\end{aligned}
	\end{equation}

Now, in order to find $\mathbf{f}_{0,0}$ that maximizes SIR, we rewrite SIR in matrix form as 
 \begin{equation}\label{equ226}
 \begin{aligned}
 \Gamma=&\frac{ \sum_{k'=0}^{K-1}\sum_{m'=0}^{M-1}{\mathbf{f}^{H}_{k',m'}\mathbf{U}_{m'}\mathbf{f}_{k',m'}}}{\sum_{k'=0}^{K-1}\sum_{m'=0}^{M-1}{\mathbf{f}^{H}_{k',m'}\mathbf{V}\mathbf{f}_{k',m'}}- {\mathbf{f}^{H}_{k',m'}\mathbf{U}_{m'}\mathbf{f}_{k',m'}}}\\&=\frac{\mathbf{f}_{0,0}^{H}\mathbf{T}_{1}\mathbf{f}_{0,0}}{\mathbf{f}_{0,0}^{H}(\mathbf{T}_{2}-\mathbf{T}_{1})\mathbf{f}_{0,0}}
 \end{aligned}
 \end{equation}
where $\mathbf{V}= \mathbf{V}^{SI}+ \mathbf{V}^{R}$ , $\mathbf{T}_{1}= \sum_{k'=0}^{K-1}\sum_{m'=0}^{M-1}\mathbf{M}^{H}_{m'M}\mathbf{S}^{H}_{k'}\mathbf{U}_{m'}\mathbf{S}_{k'}\mathbf{M}_{m'M}$ and $\mathbf{T}_{2}=\sum_{k'=0}^{K-1}\sum_{m'=0}^{M-1}\mathbf{M}^{H}_{m'M}\mathbf{S}^{H}_{k'}\mathbf{V}\mathbf{S}_{k'}\mathbf{M}_{m'M}$. Therefore, the filter optimization problem for maximizing the SIR could be formulated as
 	\begin{figure*}[t]
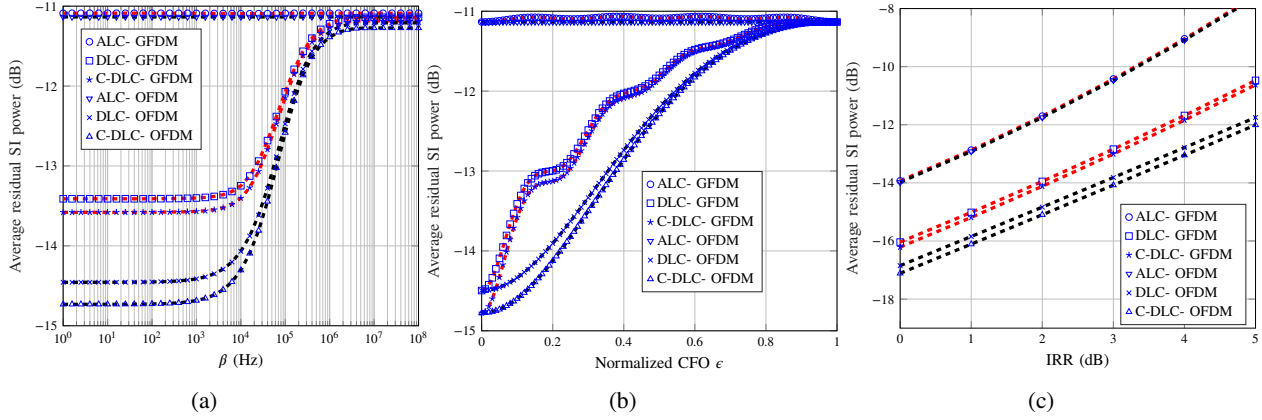

		\begin{subfigure}[b]{.3\textwidth}
		\centering
		\scalebox{.62}{\input{fig2-1.tex}}
			\caption{\fontsize{8}{8}\selectfont }
			\label{a}
		\end{subfigure} 
		\begin{subfigure}[b]{.3\textwidth}
		\centering
            \scalebox{.62}{\input{fig2-2.tex}}
			\caption{\fontsize{8}{8}\selectfont}
			\label{b}
		\end{subfigure}
				\begin{subfigure}[b]{.3\textwidth}
				\centering
            \scalebox{.62}{
%
%
\begin{tikzpicture}

\begin{axis}[%
width=3in,
height=2.7in,
at={(0.978in,0.6in)},
scale only axis,
xmin=0,
xmax=5,
xlabel style={font=\fontsize{10}{10}\selectfont},
xlabel={IRR (dB)},
ticklabel style = {font=\fontsize{8}{8}\selectfont},
ymin=-19,
ymax=-8,
ylabel style={font=\color{white!15!black}},
ylabel={Average residual SI power (dB)},
axis background/.style={fill=white},
xmajorgrids,
ymajorgrids,
legend style={font=\fontsize{9}{9}\selectfont,at={(0.62,0.01)}, anchor=south west, legend cell align=left, align=left, draw=white!15!black}
]
\addplot [only marks,color=blue, draw=none, mark size=2.0pt, mark=o, mark options={solid, blue}]
  table[row sep=crcr]{%
0	-13.9303038286095\\
1	-12.8728659210118\\
2	-11.7020516860024\\
3	-10.4222064931595\\
4	-9.0400944797033\\
5	-7.56428278056726\\
6	-6.00444786771526\\
7	-4.37070874883784\\
8	-2.67306305902156\\
9	-0.920964901384572\\
10	0.87694996127533\\
};
\addlegendentry{$\text{ALC- GFDM }$}

\addplot [color=red, dashed, line width=2.0pt,forget plot]
  table[row sep=crcr]{%
0	-13.9303038286095\\
1	-12.8728659210118\\
2	-11.7020516860024\\
3	-10.4222064931595\\
4	-9.0400944797033\\
5	-7.56428278056726\\
6	-6.00444786771526\\
7	-4.37070874883784\\
8	-2.67306305902156\\
9	-0.920964901384572\\
10	0.87694996127533\\
};

\addplot [only marks,color=blue, draw=none, mark size=2pt, mark=square, mark options={solid, blue}]
  table[row sep=crcr]{%
0	-16.0368842968875\\
1	-15.0152139263956\\
2	-13.9497100738062\\
3	-12.838940768535\\
4	-11.6806735430987\\
5	-10.4721048070187\\
6	-9.21017152757991\\
7	-7.89192463645551\\
8	-6.51492752310611\\
9	-5.0776283929507\\
10	-3.5796493729331\\
};
\addlegendentry{$\text{DLC- GFDM }$}

\addplot [color=red, dashed, line width=2.0pt,forget plot]
  table[row sep=crcr]{%
0	-16.0368842968875\\
1	-15.0152139263956\\
2	-13.9497100738062\\
3	-12.838940768535\\
4	-11.6806735430987\\
5	-10.4721048070187\\
6	-9.21017152757991\\
7	-7.89192463645551\\
8	-6.51492752310611\\
9	-5.0776283929507\\
10	-3.5796493729331\\
};

\addplot [only marks,color=blue, draw=none, mark size=2.0pt, mark=star, mark options={solid, blue}]
  table[row sep=crcr]{%
0	-16.211152040821\\
1	-15.1885967294824\\
2	-14.1204456992165\\
3	-13.0052933265013\\
4	-11.8409617022611\\
5	-10.6247434113813\\
6	-9.35372690936498\\
7	-8.0251793224291\\
8	-6.63694546449481\\
9	-5.18780884327019\\
10	-3.67775760336862\\
};
\addlegendentry{$\text{C-DLC- GFDM }$}

\addplot [color=red, dashed, line width=2.0pt,forget plot]
  table[row sep=crcr]{%
0	-16.211152040821\\
1	-15.1885967294824\\
2	-14.1204456992165\\
3	-13.0052933265013\\
4	-11.8409617022611\\
5	-10.6247434113813\\
6	-9.35372690936498\\
7	-8.0251793224291\\
8	-6.63694546449481\\
9	-5.18780884327019\\
10	-3.67775760336862\\
};

\addplot [only marks,color=blue, draw=none, mark size=2pt, mark=triangle, mark options={solid, rotate=180, blue}]
  table[row sep=crcr]{%
0	-13.9771587164975\\
1	-12.9197208088998\\
2	-11.7489065738903\\
3	-10.4690613810474\\
4	-9.08694936759127\\
5	-7.61113766845525\\
6	-6.05130275560323\\
7	-4.41756363672581\\
8	-2.71991794690955\\
9	-0.967819789272534\\
10	0.83009507338736\\
};
\addlegendentry{ALC- OFDM}

\addplot [color=black, dashed, line width=2.0pt,forget plot]
  table[row sep=crcr]{%
0	-13.9771587164975\\
1	-12.9197208088998\\
2	-11.7489065738903\\
3	-10.4690613810474\\
4	-9.08694936759127\\
5	-7.61113766845525\\
6	-6.05130275560323\\
7	-4.41756363672581\\
8	-2.71991794690955\\
9	-0.967819789272534\\
10	0.83009507338736\\
};

\addplot [only marks,color=blue, draw=none, mark size=2.0pt, mark=x, mark options={solid, blue}]
  table[row sep=crcr]{%
0	-16.848474777004\\
1	-15.8448341923573\\
2	-14.8337374106212\\
3	-13.8146537496811\\
4	-12.7866804251044\\
5	-11.7485163895893\\
6	-10.6984286591974\\
7	-9.63421483014586\\
8	-8.55316755740513\\
9	-7.4520494038452\\
10	-6.32708945662995\\
};
\addlegendentry{DLC- OFDM}

\addplot [color=black, dashed, line width=2.0pt,forget plot]
  table[row sep=crcr]{%
0	-16.848474777004\\
1	-15.8448341923573\\
2	-14.8337374106212\\
3	-13.8146537496811\\
4	-12.7866804251044\\
5	-11.7485163895893\\
6	-10.6984286591974\\
7	-9.63421483014586\\
8	-8.55316755740513\\
9	-7.4520494038452\\
10	-6.32708945662995\\
};

\addplot [only marks,color=blue, draw=none, mark size=2pt, mark=triangle, mark options={solid, blue}]
  table[row sep=crcr]{%
0	-17.1127929131381\\
1	-16.1089239772989\\
2	-15.0971324172159\\
3	-14.0768583152919\\
4	-13.04715002242\\
5	-12.0066380231997\\
6	-10.9535017869402\\
7	-9.88543383476896\\
8	-8.79960749583761\\
9	-7.6926575982105\\
10	-6.56068633957514\\
};
\addlegendentry{C-DLC- OFDM}

\addplot [color=black, dashed, line width=2.0pt,forget plot]
  table[row sep=crcr]{%
0	-17.1127929131381\\
1	-16.1089239772989\\
2	-15.0971324172159\\
3	-14.0768583152919\\
4	-13.04715002242\\
5	-12.0066380231997\\
6	-10.9535017869402\\
7	-9.88543383476896\\
8	-8.79960749583761\\
9	-7.6926575982105\\
10	-6.56068633957514\\
};

\end{axis}
\end{tikzpicture}%
}
			\caption{\fontsize{8}{8}\selectfont}
			\label{c}
		\end{subfigure}
		\caption {Average residual SI power versus 3-dB phase noise bandwidth, normalized CFO and IRR.}
		\label{fig2}
	\end{figure*}

   \begin{equation}\label{equ27}
\begin{aligned}
&\mathbf{f}^{opt}_{0,0}=\text{arg}~~ \underset{\mathbf{x}}{\text{max}}
~~ \frac{\mathbf{x}^{H}\mathbf{T}_{1}\mathbf{x}}{\mathbf{x}^{H}(\mathbf{T}_{2}-\mathbf{T}_{1})\mathbf{x}}\\
& s.t.~~ ||\mathbf{x}||^{2}=1
\end{aligned}
\end{equation}
where $\mathbf{x}\in \mathbb{C}^{MK \times 1}$ and $||\mathbf{x}||$ indicates norm of $\mathbf{x}$. Optimal receiver filter is derived by the solution that is given by \cite{filter} as
\begin{equation}\label{equ228}
\mathbf{f}^{opt}_{0,0}\propto\text{max}\left[\text{eigenvector}\left((\mathbf{T}_{2}-\mathbf{T}_{1})^{-1}\mathbf{T}_{1}\right)\right]. 
\end{equation}

Thus, we propose a receiver filter that maximizes the SIR of the FD GFDM under the RF impairments after ALC and C-DLC.      
 \section{Simulation results}
 In this section, the analytical derivations of residual SI signal power, intended signal power and SIR are verified with simulation results. Moreover,   FD GFDM and FD OFDM  are compared  the  presence of phase noise, CFO and IQ imbalance.  Finally, we optimla receiver filter with  conventional  matched filter (MF) and zero forcing (ZF) receivers. The   cyclic prefix for both  OFDM and GFDM  is  equal to the length of the channel, and the number of subcarriers is 32.    Additionally, GFDM uses  $M$=5 time slots and  root raised-cosine filter with the roll-off factor 0.1 and digital modulation 16-QAM (quadrature amplitude modulation). Sampling frequency is equal to 15.36 MHz \cite{FD-PN}. Multipath Rayleigh fading channel with total of $L=5$ taps is utilized for generating wireless channels. The power delay profile of SI channel is -30 dB, -65 dB, -70 dB and -75 dB for delays of 0, 1, 2 and 4 samples \cite{FD-PN}. Furthermore, power delay profile of desired channel  is $[-50, -75, -80, -85, -90]$ $\text{dBs}$. Same IQ imbalance level, $\text{IRR}_{Tx}=\text{IRR}_{Rx}$, is considered for the transmitter and the receiver. The  theoretical results are shown with dash lines.  
		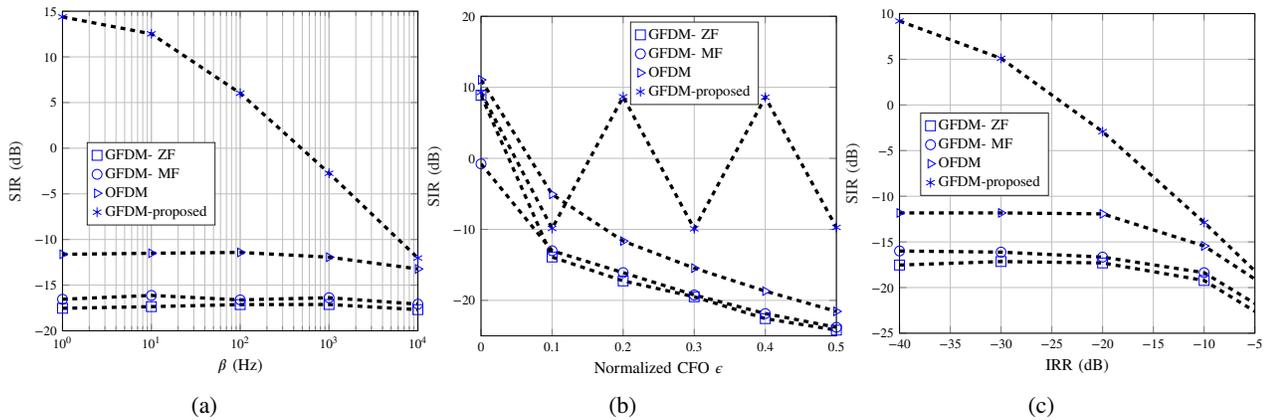
\begin{figure*}[t]
		\begin{subfigure}[b]{.3\textwidth}
		\scalebox{.62}{
%
%
\begin{tikzpicture}

\begin{axis}[%
width=3in,
height=2.7in,
at={(1.08in,0.508in)},
scale only axis,
xmode=log,
xmin=1,
xmax=10000,
xminorticks=true,
xlabel style={font=\fontsize{10}{10}\selectfont},
xlabel={$\beta$ (Hz)},
ticklabel style = {font=\fontsize{8}{8}\selectfont},
ymin=-20,
ymax=15,
ylabel style={font=\color{white!15!black}},
ylabel={SIR (dB)},
axis background/.style={fill=white},
xmajorgrids,
xminorgrids,
ymajorgrids,
legend style={font=\fontsize{9}{9}\selectfont,at={(0.43,0.59)},legend cell align=left, align=left, draw=white!15!black}
]
\addplot [only marks,color=blue, draw=none, mark size=3pt, mark=square, mark options={solid, blue}]
  table[row sep=crcr]{%
1	-17.5327720612369\\
10	-17.352224244084\\
100	-17.1275896932774\\
1000	-17.1168459705384\\
10000	-17.6796543939086\\
};
\addlegendentry{GFDM- ZF}

\addplot [color=black, dashed, line width=2.0pt,forget plot]
  table[row sep=crcr]{%
1	-17.5327720612369\\
10	-17.352224244084\\
100	-17.1275896932774\\
1000	-17.1168459705384\\
10000	-17.6796543939086\\
};

\addplot [only marks,color=blue, draw=none, mark size=3.0pt, mark=o, mark options={solid, blue}]
  table[row sep=crcr]{%
1	-16.54395932897\\
10	-16.1232824997054\\
100	-16.6121609754694\\
1000	-16.3755143625264\\
10000	-17.0754614274329\\
};
\addlegendentry{GFDM- MF}

\addplot [color=black, dashed, line width=2.0pt,forget plot]
  table[row sep=crcr]{%
1	-16.54395932897\\
10	-16.1232824997054\\
100	-16.6121609754694\\
1000	-16.3755143625264\\
10000	-17.0754614274329\\
};

\addplot [only marks,color=blue, draw=none, mark size=3pt, mark=triangle, mark options={solid, rotate=270, blue}]
  table[row sep=crcr]{%
1	-11.6211127485209\\
10	-11.5051787671978\\
100	-11.4126028728906\\
1000	-11.9333873583209\\
10000	-13.2238696466973\\
};
\addlegendentry{OFDM}

\addplot [color=black, dashed, line width=2.0pt,forget plot]
  table[row sep=crcr]{%
1	-11.6211127485209\\
10	-11.5051787671978\\
100	-11.4126028728906\\
1000	-11.9333873583209\\
10000	-13.2238696466973\\
};

\addplot [only marks,color=blue, draw=none, mark size=3.0pt, mark=asterisk, mark options={solid, blue}]
  table[row sep=crcr]{%
1	14.3744955215971\\
10	12.5213736398997\\
100	6.02404110736652\\
1000	-2.74503645595154\\
10000	-12.0235692836033\\
};
\addlegendentry{GFDM-proposed}

\addplot [color=black, dashed, line width=2.0pt,forget plot]
  table[row sep=crcr]{%
1	14.3744955215971\\
10	12.5213736398997\\
100	6.02404110736652\\
1000	-2.74503645595154\\
10000	-12.0235692836033\\
};

\end{axis}
\end{tikzpicture}%
}
			\caption{\fontsize{8}{8}\selectfont}
			\label{aa}
		\end{subfigure} 
		\begin{subfigure}[b]{.3\textwidth}
            \scalebox{.62}{
%
%
\begin{tikzpicture}

\begin{axis}[%
width=3in,
height=2.7in,
at={(1.08in,0.508in)},
scale only axis,
xmin=0,
xmax=0.5,
xlabel style={font=\fontsize{10}{10}\selectfont},
xlabel={Normalized CFO  $\epsilon$},
ticklabel style = {font=\fontsize{8}{8}\selectfont},
ymin=-25,
ymax=20,
ylabel style={font=\color{white!15!black}},
ylabel={SIR (dB)},
axis background/.style={fill=white},
xmajorgrids,
ymajorgrids,
legend style={font=\fontsize{9}{9}\selectfont,at={(0.42,0.723)}, anchor=south west, legend cell align=left, align=left, draw=white!15!black}
]
\addplot [only marks,color=blue, draw=none, mark size=3pt, mark=square, mark options={solid, blue}]
  table[row sep=crcr]{%
0	8.88374910039464\\
0.1	-13.8945279783202\\
0.2	-17.2602724612034\\
0.3	-19.4921535669957\\
0.4	-22.5677122778952\\
0.5	-24.1865507061548\\
};
\addlegendentry{GFDM- ZF}

\addplot [color=black, dashed, line width=2.0pt,forget plot]
  table[row sep=crcr]{%
0	8.88374910039464\\
0.1	-13.8945279783202\\
0.2	-17.2602724612034\\
0.3	-19.4921535669957\\
0.4	-22.5677122778952\\
0.5	-24.1865507061548\\
};

\addplot [only marks,color=blue, draw=none, mark size=3.0pt, mark=o, mark options={solid, blue}]
  table[row sep=crcr]{%
0	-0.77766502355012\\
0.1	-13.012291075994\\
0.2	-16.0935930496936\\
0.3	-19.2388849314653\\
0.4	-21.859947369946\\
0.5	-23.789867545771\\
};
\addlegendentry{GFDM- MF}

\addplot [color=black, dashed, line width=2.0pt,forget plot]
  table[row sep=crcr]{%
0	-0.77766502355012\\
0.1	-13.012291075994\\
0.2	-16.0935930496936\\
0.3	-19.2388849314653\\
0.4	-21.859947369946\\
0.5	-23.789867545771\\
};

\addplot [only marks,color=blue, draw=none, mark size=3pt, mark=triangle, mark options={solid, rotate=270, blue}]
  table[row sep=crcr]{%
0	11.0546570878443\\
0.1	-5.06456211539233\\
0.2	-11.6452595738316\\
0.3	-15.4568454830073\\
0.4	-18.6980435453802\\
0.5	-21.5522725700793\\
};
\addlegendentry{OFDM}

\addplot [color=black, dashed, line width=2.0pt,forget plot]
  table[row sep=crcr]{%
0	11.0546570878443\\
0.1	-5.06456211539233\\
0.2	-11.6452595738316\\
0.3	-15.4568454830073\\
0.4	-18.6980435453802\\
0.5	-21.5522725700793\\
};

\addplot [only marks,color=blue, draw=none, mark size=3.0pt, mark=asterisk, mark options={solid, blue}]
  table[row sep=crcr]{%
0	9.27678793683199\\
0.1	-9.87563482271843\\
0.2	8.66984315854217\\
0.3	-9.92306106580506\\
0.4	8.60014708220822\\
0.5	-9.72809683476153\\
};
\addlegendentry{GFDM-proposed}

\addplot [color=black, dashed, line width=2.0pt,forget plot]
  table[row sep=crcr]{%
0	9.27678793683199\\
0.1	-9.87563482271843\\
0.2	8.66984315854217\\
0.3	-9.92306106580506\\
0.4	8.60014708220822\\
0.5	-9.72809683476153\\
};

\end{axis}
\end{tikzpicture}%
}
			\caption{\fontsize{8}{8}\selectfont}
			\label{bb}
		\end{subfigure}
				\begin{subfigure}[b]{.3\textwidth}
            \scalebox{.62}{
%
%
\begin{tikzpicture}

\begin{axis}[%
width=3in,
height=2.7in,
at={(1.08in,0.508in)},
scale only axis,
xmin=-40,
xmax=-5,
xlabel style={font=\fontsize{10}{10}\selectfont},
xlabel={IRR (dB)},
ticklabel style = {font=\fontsize{8}{8}\selectfont},
ymin=-25,
ymax=10,
ylabel style={font=\color{white!15!black}},
ylabel={SIR (dB)},
axis background/.style={fill=white},
xmajorgrids,
ymajorgrids,
legend style={font=\fontsize{9}{9}\selectfont,at={(0.42,0.69)},legend cell align=left, align=left, draw=white!15!black}
]
\addplot [only marks,color=blue, draw=none, mark size=3pt, mark=square, mark options={solid, blue}]
  table[row sep=crcr]{%
-40	-17.5326827108384\\
-30	-17.1507158828589\\
-20	-17.3038252107453\\
-10	-19.2228242538163\\
0	-25.98522444021\\
10	-38.973662879283\\
20	-57.1252845306932\\
};
\addlegendentry{GFDM- ZF}

\addplot [color=black, dashed, line width=2.0pt,forget plot]
  table[row sep=crcr]{%
-40	-17.5326827108384\\
-30	-17.1507158828589\\
-20	-17.3038252107453\\
-10	-19.2228242538163\\
0	-25.98522444021\\
10	-38.973662879283\\
20	-57.1252845306932\\
};

\addplot [only marks,color=blue, draw=none, mark size=3.0pt, mark=o, mark options={solid, blue}]
  table[row sep=crcr]{%
-40	-15.9953758059624\\
-30	-16.118553813435\\
-20	-16.6620307136968\\
-10	-18.36975772431\\
0	-25.1181177369688\\
10	-38.5157243911658\\
20	-56.5768603929037\\
};
\addlegendentry{GFDM- MF}

\addplot [color=black, dashed, line width=2.0pt,forget plot]
  table[row sep=crcr]{%
-40	-15.9953758059624\\
-30	-16.118553813435\\
-20	-16.6620307136968\\
-10	-18.36975772431\\
0	-25.1181177369688\\
10	-38.5157243911658\\
20	-56.5768603929037\\
};

\addplot [only marks,color=blue, draw=none, mark size=3pt, mark=triangle, mark options={solid, rotate=270, blue}]
  table[row sep=crcr]{%
-40	-11.8265656701863\\
-30	-11.8249835092794\\
-20	-11.9298367745526\\
-10	-15.4423458278616\\
0	-22.6993244967425\\
10	-35.2787916465\\
20	-52.2716621040708\\
};
\addlegendentry{OFDM}

\addplot [color=black, dashed, line width=2.0pt,forget plot]
  table[row sep=crcr]{%
-40	-11.8265656701863\\
-30	-11.8249835092794\\
-20	-11.9298367745526\\
-10	-15.4423458278616\\
0	-22.6993244967425\\
10	-35.2787916465\\
20	-52.2716621040708\\
};

\addplot [only marks,color=blue, draw=none, mark size=3.0pt, mark=asterisk, mark options={solid, blue}]
  table[row sep=crcr]{%
-40	9.1880219473146\\
-30	5.1016067047536\\
-20	-2.92261016616741\\
-10	-12.8716861285289\\
0	-23.4814994729873\\
10	-35.8914848748708\\
20	-47.002654047938\\
};
\addlegendentry{GFDM-proposed}

\addplot [color=black, dashed, line width=2.0pt,forget plot]
  table[row sep=crcr]{%
-40	9.1880219473146\\
-30	5.1016067047536\\
-20	-2.92261016616741\\
-10	-12.8716861285289\\
0	-23.4814994729873\\
10	-35.8914848748708\\
20	-47.002654047938\\
};

\end{axis}
\end{tikzpicture}%
}
			\caption{\fontsize{8}{8}\selectfont}
			\label{cc}
		\end{subfigure}
		\caption {SIR versus 3-dB phase noise bandwidth, normalized CFO and IRR.}
		\label{fig4}
	\end{figure*}

 In  Fig. \ref{fig2},   the residual SI power is plotted versus the 3-dB phase noise bandwidth, normalized CFO and IRR when ZF receiver is utilized for GFDM. For illustrating the impact of the conjugate residual SI cancellation, we consider both classical DLC (legend DLC) and C-DLC. As can be seen, the simulation results fully match the derived  residual SI power. Moreover, C-DLC lowers  SI power compared to  DLC. Thus, cancelling the conjugate residual SI improves the performance significantly. In Fig. \ref{a}, $\epsilon=0.1$ and $\text{IRR}=2.5$\;dB, residual SI is plotted as a function of $\beta$.  With  increasing 3-dB phase noise bandwidth, post-DLC and post-C-DLC, average residual SI increases and  saturates at a constant value that is lower than post-ALC residual SI.  Furthermore, in Fig. \ref{b}, $\beta=10$\;Hz and $\text{IRR}=2.5$\;dB and $\epsilon$ is changed. By increasing $\epsilon$, average residual SI power after DLC boosts and approaches to average residual SI power after ALC. According to  (\ref{equ12})-(\ref{equ15}), 3-dB phase noise bandwidth, $\beta$, and normalized CFO, $\epsilon$, appear in the  exponential terms,  and  the trends in  Fig. \ref{a} and  Fig. \ref{b} for higher values of $\beta$ and $\epsilon$  can be  due to their appearance in the exponential function.  The residual SI  increases with $\text{IRR}$ (Fig. \ref{c}). In this figure, $\beta=10$\;Hz and $\epsilon=0.1$ are fixed parameters and $\text{IRR}$ is the independent variable. Finally,  FD OFDM outperforms FD GFDM  because of non-orthogonality of GFDM that generates more interference terms.  

 Fig. \ref{fig4} depicts  SIR as a function of  3-dB phase noise bandwidth, normalized CFO and IRR between FD GFDM with MF, ZF, proposed receiver filters and FD OFDM. Perfect  match between the theoretical SIR (\ref{equ26}) and numerical simulations can be observed.  Moreover,   FD OFDM  achieves larger SIR  than FD GFDM with conventional filters, emphasizing  the necessity of designing an optimal receiver filter. Fig. \ref{aa} illustrates SIR versus $\beta$ when  $\epsilon=0.2$ and $\text{IRR}=-37.5$\;dB. Obviously, by increasing $\beta$, SIR decreases in all cases. We can see that the SIR achieved by the optimized filter exceeds that of the others, e.g. in $\beta=10$\;Hz, SIR of the proposed filter is 25\;dB higher than FD OFDM. Fig. \ref{bb} plots  SIR as a function of $\epsilon$ when $\beta=50$\;Hz and $\text{IRR}=-37.5$\;dB. In FD GFDM  with conventional filters and OFDM, higher value of CFO,  $\epsilon$, lowers  SIR, which is not surprising because  frequency offset results in interference in general.  Furthermore, the optimum filter with GFDM   outperforms  the other options, e.g. for  $\epsilon=0.2$,  it achieves 20 dB higher SIR than  FD OFDM. Fig. \ref{cc} represents the SIR versus IRR when $\beta=50$\;Hz and $\epsilon=0.2$. In all cases, higher values of the IRR provides lower SIR. Furthermore, FD  GFDM with the proposed filter is always  better than FD OFDM, e.g. in $\text{IRR}=-30 $\;dB, the  gap is  17\;dB, indicating  a significant improvement.   

\section{Conclusion}\label{sec:conclusion}

In this paper, we investigated an  FD GFDM transceiver in the  presence of three  RF impairments, namely phase noise, CFO and IQ imbalance. We considered both  analog and digital SI cancellations and developed a  complementary digital  SI suppression method. Closed-form expressions for residual SI power and desired signal power were derived. The  receiver filter for maximizing the SIR were proposed. Simulation results verified  the analytical derivations. We observed that RF impairments  degrade the SI cancellation methods. Moreover, the SIR results for FD GFDM with MF, ZF and proposed filter and FD OFDM show that proposed filter outperforms the other options.  For instance, FD GFDM with the designed receiver filter achieves 25 \;dB higher SIR  than  FD OFDM when $\beta=10~\text{Hz}$, $\epsilon=0.2$ and $\text{IRR}= -37.5$\;dB.    

\bibliographystyle{IEEE}
\bibliography{IEEEabrv,thesis-bib}
\end{document}